\documentclass[aps,prd,preprintnumbers,twocolumn,groupedaddress,nofootinbib]{revtex4}
\usepackage{graphicx}
\usepackage{latexsym}
\def\beq{\begin{equation}}
\def\eeq{\end{equation}}
\def\bey{\begin{eqnarray}}
\def\eey{\end{eqnarray}}

\def\lsim{\mathrel{\raise.3ex\hbox{$<$\kern-.75em\lower1ex\hbox{$\sim$}}}}
\def\gsim{\mathrel{\raise.3ex\hbox{$>$\kern-.75em\lower1ex\hbox{$\sim$}}}}

\begin{document}

\title{Are There Hints of Light Stops in Recent Higgs Search Results?}  
\author{Matthew R.~Buckley$^1$ and Dan Hooper$^{1,2}$}
\affiliation{$^1$Center for Particle Astrophysics, Fermi National Accelerator Laboratory, Batavia, IL 60510, USA}
\affiliation{$^2$Department of Astronomy and Astrophysics, University of Chicago, Chicago, IL 60637, USA}

\date{\today}

\begin{abstract}
The recent discovery at the LHC by the CMS and ATLAS collaborations of the Higgs boson presents, at long last, direct probes of the mechanism for electroweak symmetry breaking. While it is clear from the observations that the new particle plays some role in this process, it is not yet apparent whether the couplings and widths of the observed particle match those predicted by the Standard Model. In this paper, we perform a global fit of the Higgs results from the LHC and Tevatron. While these results could be subject to as-yet-unknown systematics, we find that the data are significantly better fit by a Higgs with a suppressed width to gluon-gluon and an enhanced width to $\gamma\gamma$, relative to the predictions of the Standard Model. After considering a variety of new physics scenarios which could potenially modify these widths, we find that the most promising possibility is the addition of a new colored, charged particle, with a large coupling to the Higgs. Of particular interest is a light, and highly mixed, stop, which we show can provide the required alterations to the combination of $gg$ and $\gamma\gamma$ widths.
\end{abstract}

\pacs{95.35.+d; FERMILAB-PUB-12-351-A}
\maketitle

\section{Introduction}

After almost five decades since being hypothesized \cite{Englert:1964et}, experimental evidence for the Higgs boson has appeared. The Large Hadron Collider's (LHC) ATLAS and CMS collaborations have reported discovery of an even-integer-spin particle with production and decay modes approximately consistent with those of the Standard Model Higgs. ATLAS \cite{ATLASdiscovery} has reported discovery of such a particle with a best fit mass of $126.5$~GeV at $5.0\sigma$, while CMS \cite{CMSdiscovery} finds a particle with a mass of $125.3\pm0.6$~GeV with $4.9\sigma$ significance. Furthermore, the Tevatron's CDF and D\O\ Collaborations have each observed a statistically significant excess among events in their $b\bar{b}+W/Z$ channels \cite{TevatronBB,TEVNPH:2012ab,CDF:2012cn}, corresponding to a rate that is roughly consistent with that predicted from the production and decay of a 125 GeV Standard Model Higgs boson. The current body of evidence strongly supports the conclusion that these experiments are observing the scalar particle that is responsible for electroweak symmetry breaking.

The quest for the Higgs, however, does not end with discovery. Increasingly precise measurements of this particle's production cross sections, decay widths, and mass can provide potentially valuable probes of physics beyond the Standard Model. For example, within many extensions of the Standard Model (most notably supersymmetry \cite{Dimopoulos:1981zb}) there is not one Higgs boson but several. These Higgs bosons can mix among each other, leading to modified couplings relative to those predicted by the Standard Model. Furthermore, in the Standard Model and extensions thereof, the Higgs' effective couplings to gluons and photons are induced only at the loop-level \cite{Ellis:1975ap} and thus can be potentially sensitive to the presence of new charged or colored particles with significant couplings to the Higgs. In such scenarios, the particle that is currently being observed at the LHC and Tevatron may be similar to the Higgs boson of the Standard Model, but with modfied couplings. Precision measurements of the production rates and branching fractions of this particle could thus provide a new window through which to potentially reveal new physics. For recent studies of 2011 Tevatron and LHC results within the context of a supersymmetric Higgs sector, see Refs.~\cite{Kane:2011kj,Hall:2011aa,Baer:2011ab,Feng:2011aa,Heinemeyer:2011aa,Arbey:2011ab,Draper:2011aa, Carena:2011aa,Ellwanger:2011aa,Akula:2011aa,Kadastik:2011aa,Gunion:2012zd,King:2012is,Cao:2012fz,Aparicio:2012iw,Christensen:2012ei,Ajaib:2012vc,Brummer:2012ns}.

In this article, we study the rates observed by the LHC and Tevatron experiments in various Higgs search channels and compare them to those predicted for a Standard Model Higgs boson. We allow the ratios of widths to Standard Model particles to vary, and consider the effect this has on the global fit to the ATLAS, CMS and CDF/D\O\ observations across all channels. As had been noted in a number of similar studies of previous LHC and Tevatron data~\cite{Giardino:2012ww,Carena:2012xa,Carena:2012gp,Blum:2012ii,Bonne:2012im,Chang:2012tb,Carmi:2012zd,Azatov:2012ga,Wang:2012gm,Azatov:2012wq,Akeroyd:2012ms,Espinosa:2012vu,Arhrib:2012yv,Barroso:2012wz,Bellazzini:2012tv,Klute:2012pu,Draper:2012xt,Gabrielli:2012hd,Ellis:2012rx}, these observations appear to favor an increased rate of $h\to\gamma\gamma$, as well as reduced production cross sections in the channels that rely on $gg \to h$, as compared to the Standard Model predictions. Incorporating all publicly available data, we quantify this preference, and find that, neglecting possible systematic effects, increasing $\Gamma(h\to \gamma \gamma)$ by a factor of approximately three while decreasing $\Gamma(h \to g g)$ by a factor of two very significantly improves the quality of the global fit, reducing the $\chi^2$ by 9.9.

The remainder of this paper is structured as follows. In Sec.~\ref{datasec}, we summarize the current status of Higgs searches at the LHC and Tevatron and compare the observed rates to those predicted for a Standard Model Higgs boson. In Sec.~\ref{widthcouplings}, we use these data to constrain the decay widths and effective couplings of the particle being observed. In Sec.~\ref{scenarios}, we discuss classes of beyond the Standard Model physics scenarios which could help to alleviate the tension between the observed and predicted rates. We find that the statistical tension between the theory and experiment can be significantly relieved by the addition of new particles with both color and electric charge, and large couplings to the Higgs. An obvious realization of such a particle is the stop squark of supersymmetry.  We find that a light ($m_{\tilde{t_1}} \lsim 300$ GeV) and highly mixed stop can provide a good fit to these observations.

\section{Higgs Boson Searches At The LHC and Tevatron}
\label{datasec}

Standard Model Higgs production at both the LHC and the Tevatron is dominated by gluon-gluon fusion, made possible by the effective coupling induced by a top quark loop \cite{Spira:1995rr,Anastasiou:2008tj,deFlorian:2009hc}. Smaller but not insignificant rates for Higgs production also occur through vector boson fusion (VBF) and through Higgs production in association with a $W$ or $Z$. At $\sqrt{s}=7(8)$~TeV, the cross-sections for these channels are \cite{Dittmaier:2012vm}:
\begin{eqnarray}
\sigma(g g \to h) & = & 15.3 \pm 2.3~(19.5\pm 2.9)~\mbox{pb}, \\
\sigma(p p \to j j h) & = & 1.21 \pm 0.03~(1.56\pm0.05)~\mbox{pb}, \nonumber \\
\sigma(p p\to Wh) & = & 0.57 \pm 0.02~(0.70\pm0.03)~\mbox{pb}, \nonumber \\
\sigma(p p\to Zh) & = & 0.32\pm0.02~(0.39\pm0.02)~\mbox{pb}. \nonumber
\end{eqnarray}
While the cross sections for the later three processes are much smaller than those from gluon-gluon fusion, the additional very forward jets present in VBF events and the high transverse momentum typical among associated production events makes these channels important for LHC Higgs studies, especially in the case of $h \rightarrow \gamma \gamma$.

For a mass of $m_h \approx 125$ GeV, as reported by the high resolution $\gamma \gamma$ and $ZZ\to 4\ell$ channels at CMS and ATLAS \cite{ATLASdiscovery,CMSdiscovery}, the Standard Model Higgs boson is predicted to decay primarily to $b\bar{b}$ (58\%) and $W^+W^-$ (22\%), with smaller branching fractions to $gg$ (8.5\%), $\tau^+ \tau^-$ (6.4\%), $c\bar{c}$ (2.7\%), $ZZ$ (2.7\%), and $\gamma \gamma$ (0.22\%) \cite{Dittmaier:2012vm}. Higgs decays to $gg$ are completely invisible at hadron colliders, due to the very large QCD multi-jet background. However, as the loop-induced coupling responsible for this branching ratio also sets the $gg\to h$ cross-section, it is of critical importance to LHC and Tevatron Higgs phenomenology.
 
Higgs searches have been conducted in many channels by the LHC and Tevatron experimental collaborations. Some Higgs decay modes (such as $h\rightarrow W^+W^-$, $ZZ$, and $\tau^+\tau^-$) yield signatures that are significantly distinctive to study independently of the production mechanism. Others (such as $h\rightarrow b\bar{b}$) can only be identified above backgrounds when observed in conjunction with an additional gauge boson, and thus rely on those events in which the Higgs is produced in association with an additional $W$ or $Z$ \cite{CMSbb}.

\begin{figure*}[t]
\centering
%\vspace{-1.0cm}
\includegraphics[angle=0.0,width=6.0in]{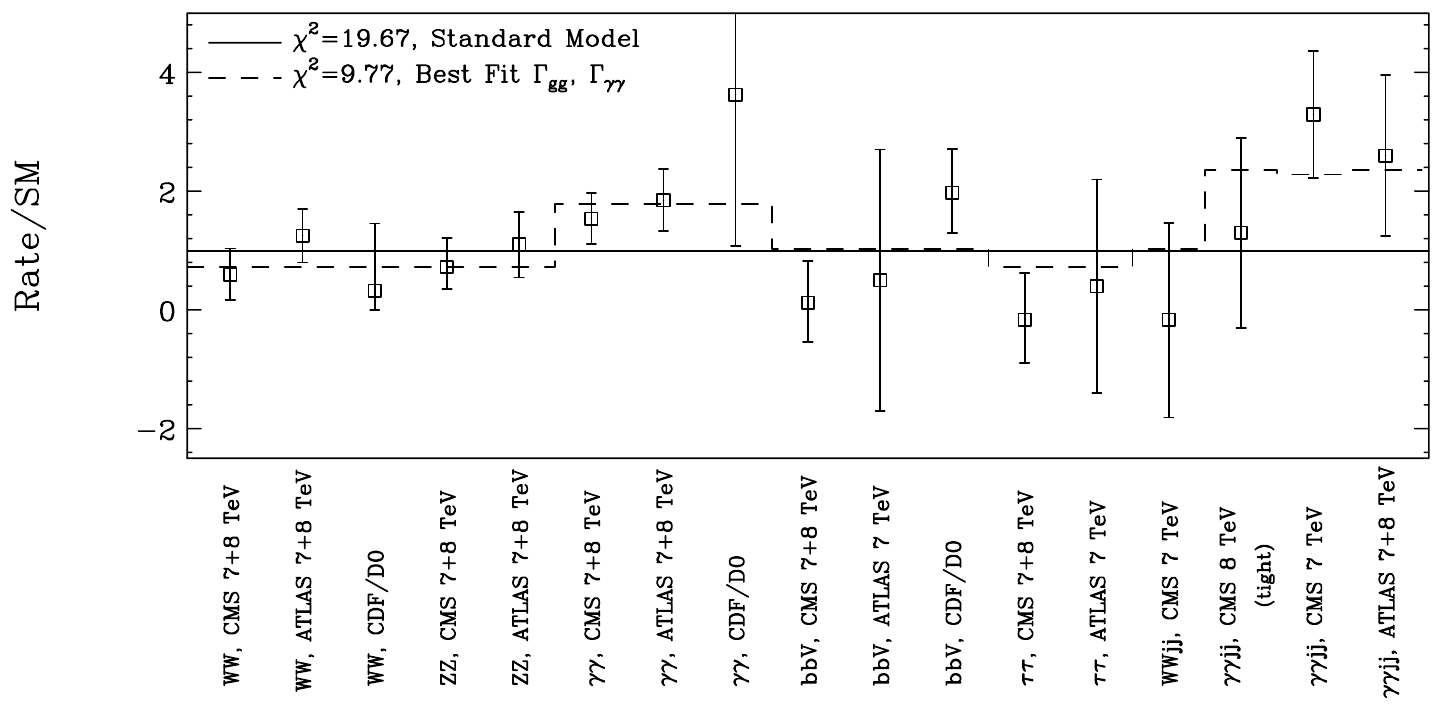}
\caption{The observed rates in various Higgs search channels at the LHC and Tevatron, compared to those predicted from a 125 GeV Standard Model Higgs boson. The Standard Model Higgs boson provides a somewhat poor global fit ($\chi^2=19.67$, over 17-1 degrees-of-freedom). If the Higgs' decay widths to gluons and photons are modified to their best fit values (0.66 and 2.5 times the Standard Model values, respectively), the global fit improves significantly to $\chi^2=9.77$, over 17-3 degrees-of-freedom.}
\label{data}
\end{figure*}

\begin{figure}[!]
\centering
%\vspace{-1.0cm}
\includegraphics[angle=0.0,width=3.4in]{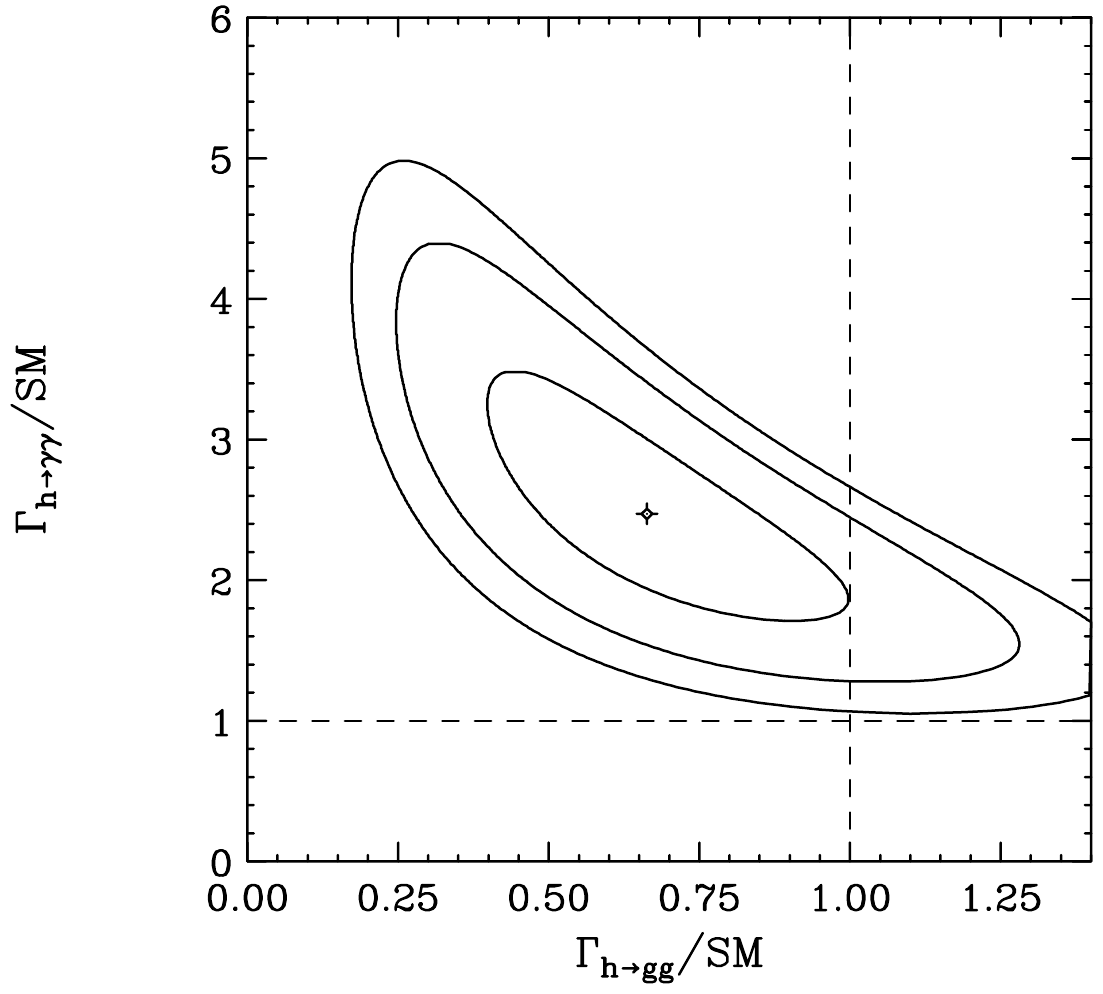}
\caption{The 68\%, 95\% and 99\% confidence level contours of the global fit to the data summarized in Fig.~\ref{data}, allowing the decay widths of the Higgs to photons and gluons to vary while keeping all others fixed. We have also fixed the mass of the Higgs to 125 GeV. The best fit point (shown as a cross) yields an overall $\chi^2$ of 9.77, over 17-3 degrees-of-freedom. In contrast, the Standard Model case, located at (1,1) in this figure, yields $\chi^2=19.67$, and at face value appears to be disfavored over the best-fit case at the 99\% confidence level.}
\label{fitwidth}
\end{figure}

Further search strategies focus on events which primarily result from VBF (such as the $pp\rightarrow jjh \rightarrow jjWW$ channel studied by the CMS Collaboration \cite{CMSvbfWW}). ATLAS and CMS have applied a number of different sets of cuts on their searches for $h \rightarrow \gamma \gamma$ \cite{ATLASdiscovery,CMSdiscovery,Chatrchyan:2012tw,ATLAS:2012ad,CMSvbf}, which have the effect of focusing on and isolating different Higgs production mechanisms. However, while the event selection cuts designed to zero in on VBF events are very effective, the large ratio of cross sections means that even a small survival probability of gluon-gluon fusion events can have large consequences. Using MadGraph5~\cite{Alwall:2011uj} combined with Pythia6~\cite{Sjostrand:2006za} to simulate Higgs production and PGS4~\cite{PGS} for detector simulation, we match the $\sim 80\%$ purity of the $p p \rightarrow \gamma\gamma j j$ ``di-jet tight'' selection criteria as reported by the CMS experiment \cite{CMSvbf} and find that (at $\sqrt{s} = 8$~TeV) it is sensitive to the following approximate combination:
\begin{equation}
\left[0.02 \, \sigma(gg\to h)+\sigma(pp\to j j h) \right] \times \mbox{BR}(h\to \gamma\gamma).
\end{equation}
The ATLAS $p p \to hX \to \gamma\gamma X$ \cite{ATLAS:2012ad} originally reported in the July 4, 2012 announcement has been updated \cite{ATLASVBF} to a two-photon plus two-jet search which isolates a combination of cross sections similar to that the CMS result. These simulations agree well with the results found in Ref~\cite{Giardino:2012ww}. Note that we do not include errors on the admixture of gluon fusion and VBF contributions when performing our global fits in Sec.~\ref{widthcouplings}.

In Fig.~\ref{data} we present a summary of the Higgs search results, shown as a comparison between the observed rate in a given channel and the rate predicted for a 125 GeV Standard Model Higgs boson, updated using the most recent ATLAS results \cite{ATLASupdate}. Standing out among these results is that all six measurements of channels involving $h\rightarrow \gamma \gamma$ have been observed at higher rates than predicted, possibly suggesting an enhanced Higgs decay width to photons. It is also the case that the rates observed in six out of the seven measurements dominated by gluon-gluon fusion ($h \rightarrow W^+ W^-$, $ZZ$, $\tau^+ \tau^-$) are lower than predicted, possibly favoring a reduced Higgs width to gluons. We find that the Standard Model Higgs without any modified widths provides a somewhat poor fit to the combined data ($\chi^2=19.67$, over 17-1 degrees-of-freedom). In the following section, we will consider variations to the Higgs' widths and couplings and discuss how the quality of this fit might be improved.

%%%%%%%%

%On July 4, 2012, the ATLAS and CMS collaborations reported the results of their Higgs analysis of XXX fb$^-1$ of data, taken at a center-of-mass energy of 8 TeV. These results included searches for Higgs bosons in the XXX, XXX... channels. 

%Two days earlier, the CDF and D\O\ Collaborations updated the results of their Higgs searches, reporting a modest excess with a local significance of 3.0$\sigma$ (or 2.5$\sigma$ including the look elsewhere effect). The bulk of this signal consists of Higgs production in association with a gauge boson, followed by Higgs decay to $b\bar{b}$. The Tevatron's significance in this channel channel alone was reported to be 3.2$\sigma$ (2.9$\sigma$ with look elsewhere)~\cite{CDF:2012cn}
%. These results supersede previous and similar reports from CDF and D\O\ ~\cite{TevatronBB,TEVNPH:2012ab}.

%DESCRIBE HOW WE GO FROM OBSERVATIONS TO WIDTHS...

\section{Constraining The Decay Widths and Couplings of the Higgs}
\label{widthcouplings}

In this section, we consider departures from the Higgs widths and couplings predicted by the Standard Model, and discuss how such variations impact the global fit to the data shown in Fig.~\ref{data}. As an initial point of departure, we focus on modifications to the Higgs decay widths to photons and gluons. As these couplings are only induced at loop-level, contributions from new physics are more likely to significantly influence these channels than those which occur through tree-level couplings.

In Fig.~\ref{fitwidth}, we show the results of a global fit to the data summarized in Fig.~\ref{data}. We have allowed the decay widths of the Higgs to $\gamma \gamma$ and $gg$ to vary, while holding all other decay widths fixed to the Standard Model prediction. We find that the best fit occurs for values of $\Gamma_{h \rightarrow \gamma\gamma}$ and $\Gamma_{h \rightarrow gg}$ that are approximately 2.5 and 0.66 times their Standard Model values, yielding an overall fit of $\chi^2=9.77$ over 17-3 degrees-of-freedom. Modifying these two loop-induced decay widths can potentially improve the global fit by $\Delta \chi^2 = 9.9$ relative to the Standard Model, while only adding two additional degrees-of-freedom. If this data is taken at face value, without consideration of any unknown systematics or other uncertainties, this fit favors modified values of $\Gamma_{h \rightarrow \gamma\gamma}$ and $\Gamma_{h \rightarrow gg}$ over those of the Standard Model at the 99\% confidence level. 

Taking this approach a step further, we have performed a global fit varying not only $\Gamma_{h \rightarrow \gamma\gamma}$ and $\Gamma_{h \rightarrow gg}$, but also the Higgs decay widths to $W^+W^-$/$ZZ$, $b\bar{b}$, and $\tau^+ \tau^-$.\footnote{As custodial $SU(2)$ forces us to require $\Gamma_{h \rightarrow WW}/\Gamma^{\rm SM}_{h \rightarrow WW}=\Gamma_{h \rightarrow ZZ}/\Gamma^{\rm SM}_{h \rightarrow ZZ}$, we chose to not vary these two widths independently of each other.} The best fit set of values found in this scan further improves the global fit to $\chi^2=8.49$. As the addition of these three new free parameters only improves the fit beyond the previous exercise by $\Delta \chi^2=1.3$, there does not appear to be any support for modifications of the tree-level couplings of the Higgs. We have also considered the possibility that the Higgs has a significant decay width to invisible final states, but find that this does not improve the global fit.

\section{Signs of Light Stops?}
\label{scenarios}

In this section, we discuss some of the types of physics beyond the Standard Model that may lead to improvements to the global fit, as discussed above. As outlined in the previous section, we focus on new particles which modify the loop-induced Higgs widths to photons and gluons. In general, new charged particles can lead to modifications to $\Gamma_{h\rightarrow \gamma\gamma}$ while new colored particles can alter $\Gamma_{h\rightarrow gg}$.

\begin{figure}[t]
\centering
%\vspace{-1.0cm}
\includegraphics[angle=0.0,width=3.4in]{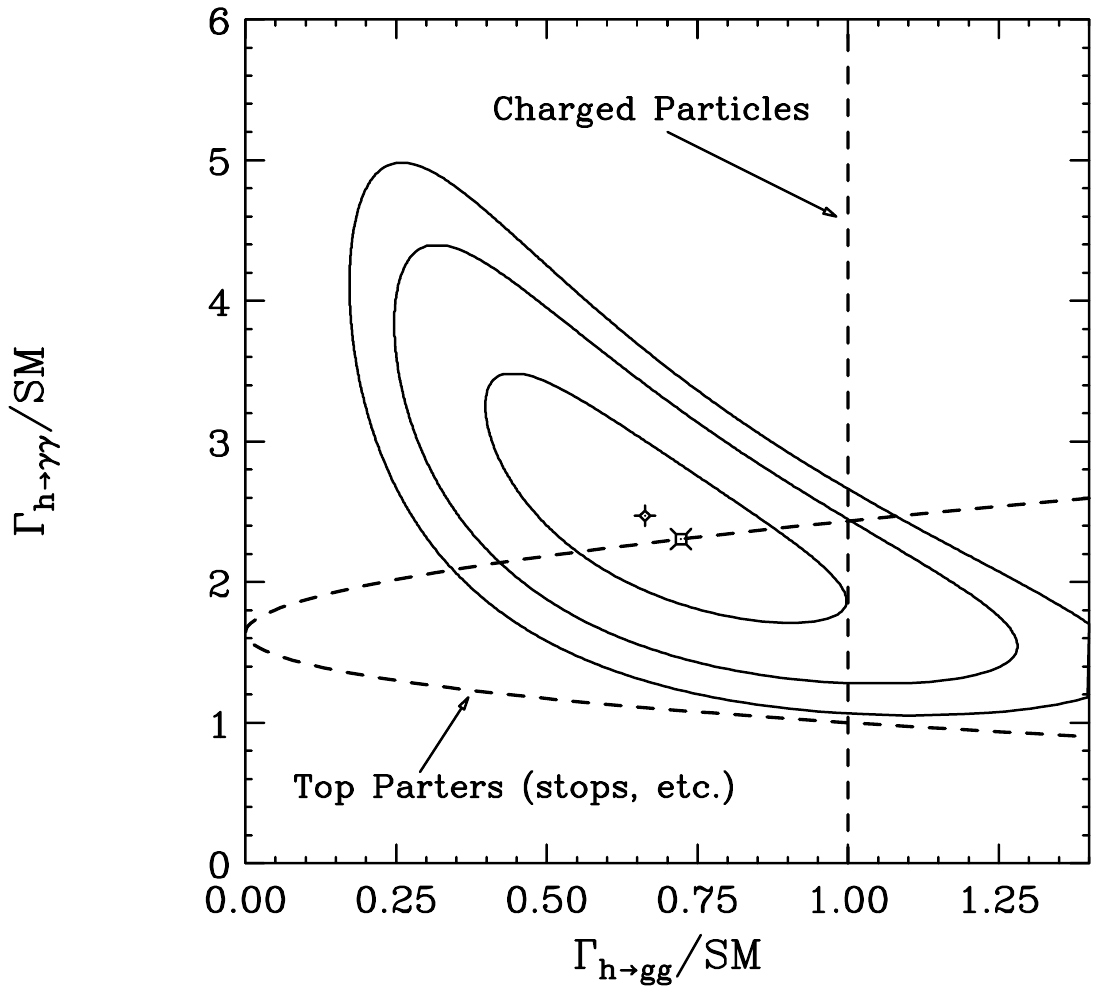}
\caption{The 68\%, 95\% and 99\% confidence level contours of the global fit to the data (as shown in Fig.~\ref{fitwidth}), compared to the range of decay widths of the Higgs that can result from the addition of new charged particles, or new charged and colored particles (top partners, defined as particles with Higgs coupling, charge, and color equal to that of the top quark).}
\label{contours}
\end{figure}

\begin{figure}[t]
\centering
%\vspace{-1.0cm}
\includegraphics[angle=0.0,width=3.4in]{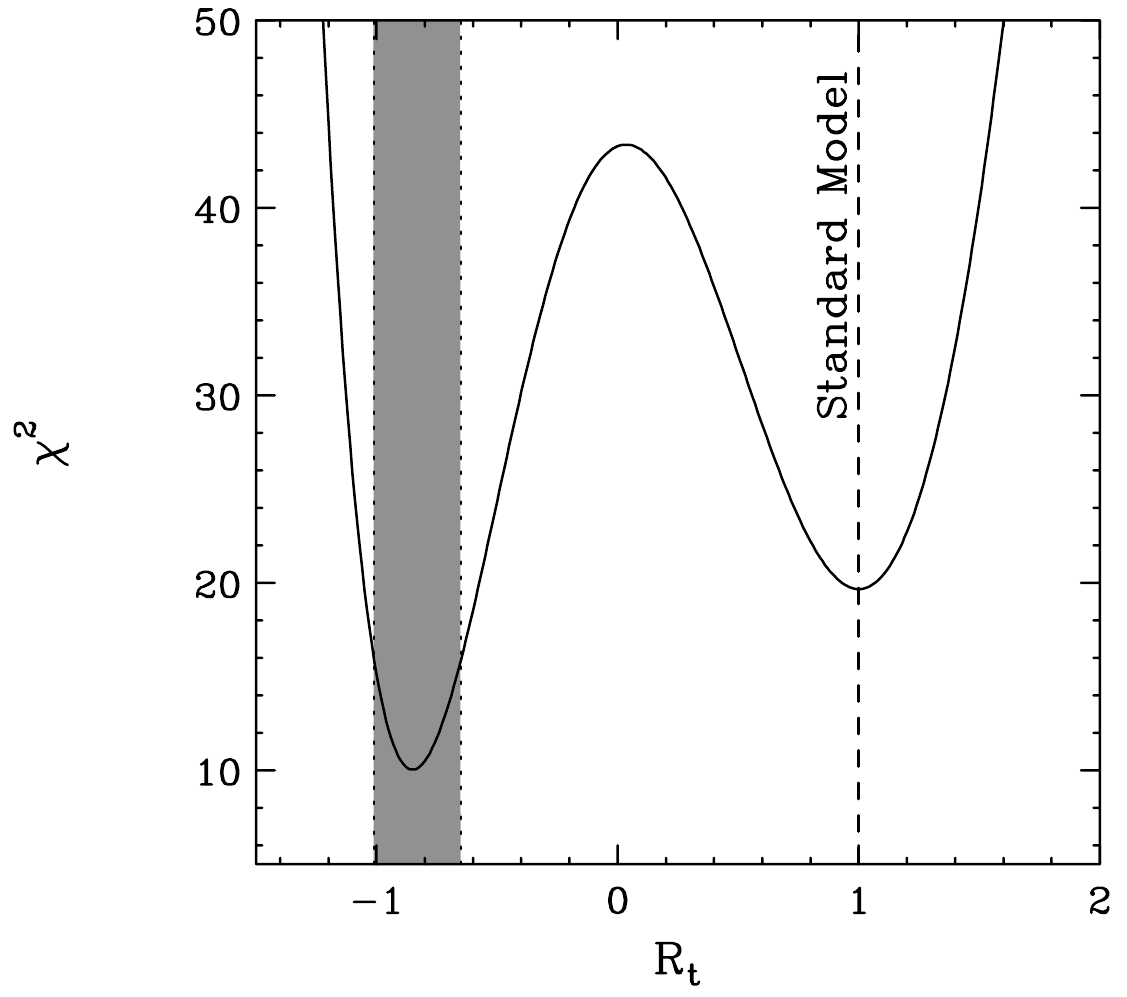}
\caption{The quality of the global fit as a function of the parameter $R_t \equiv y_t/y^{\rm SM}_t$. The miniumum near $R_t = -0.85$ (corresponding to the point marked by a star on Fig.~\ref{contours}) provides a significantly better fit to the data than the Standard Model case.}
\label{Rt}
\end{figure}

In Fig.~\ref{contours}, we show how classes of new particles impact these decay widths. Additional charged particles simply alter the width to photons, moving vertically in the plane of the figure. Particles with both charge and QCD color, however, alter both of these widths simultaneously. As a concrete and well motivated class of examples, we consider generic partners of the top quark (particles with Higgs coupling, charge, and color equal to that of the top quark). As can be seen in the figure, the inclusion of a top partner particle can potentially lead to a good fit to the data, falling within the approximate 68\% confidence region.

Specializing to a top partner, such a particle's effect on the Higgs decay widths can be parameterized by its modification of the effective Higgs-top-top coupling: $R_t \equiv y_t/y^{\rm SM}_t$ (see, for example, Ref.~\cite{Giardino:2012ww}). In Fig.~\ref{Rt}, we show how the quality of the global fit changes with this parameter. The best fit is found for $R_t \simeq -0.85$, with $\chi^2=10.03$ over 17-2 degrees-of-freedom (corresponding to the point marked with a star shown along the top partner contour in Fig.~\ref{contours}). Once again, this marks a considerable improvement over the fit to the Standard Model ($\Delta \chi^2 =9.6$). Also shown as a shaded region in Fig.~\ref{Rt} is the 95\% confidence region around this best fit value of $R_t$.

The best known and most well motivated example of a top partner are the supersymmetric partners of the top quark, known as stops. The presence of stops modifies $R_t$ at leading order~\cite{Blum:2012ii} from its Standard Model value of one as follows:
\begin{equation}
R_t \approx 1+ \frac{m^2_t}{4} \bigg[\frac{1}{m^2_{\tilde{t}_1}}+\frac{1}{m^2_{\tilde{t}_2}}-\frac{(A_t-\mu \cot \beta)^2}{m^2_{\tilde{t}_1} m^2_{\tilde{t}_2}}\bigg],
\end{equation}
where $m_{\tilde{t}_{1,2}}$ are the masses of the two stops, $A_t$ is the top trilinear coupling, $\mu$ is the higgsino mass parameter, and $\tan \beta$ is the ratio of the vacuum expectation values of the two higgs doublets of the MSSM. To achieve the negative values of $R_t$ as are favored by our fit, we must require that the mixing term dominates. In this limit, and for $A_t \gg \mu \cot \beta$, this reduces to $R_t \sim 1 - (m^2_t A^2_t /4 m^2_{\tilde{t}_1} m^2_{\tilde{t}_2})$. Thus to achieve the desired values of $R_t$, we are required to consider light and highly mixed stops.

\begin{figure}[t]
\centering
%\vspace{-1.0cm}
\includegraphics[angle=0.0,width=3.4in]{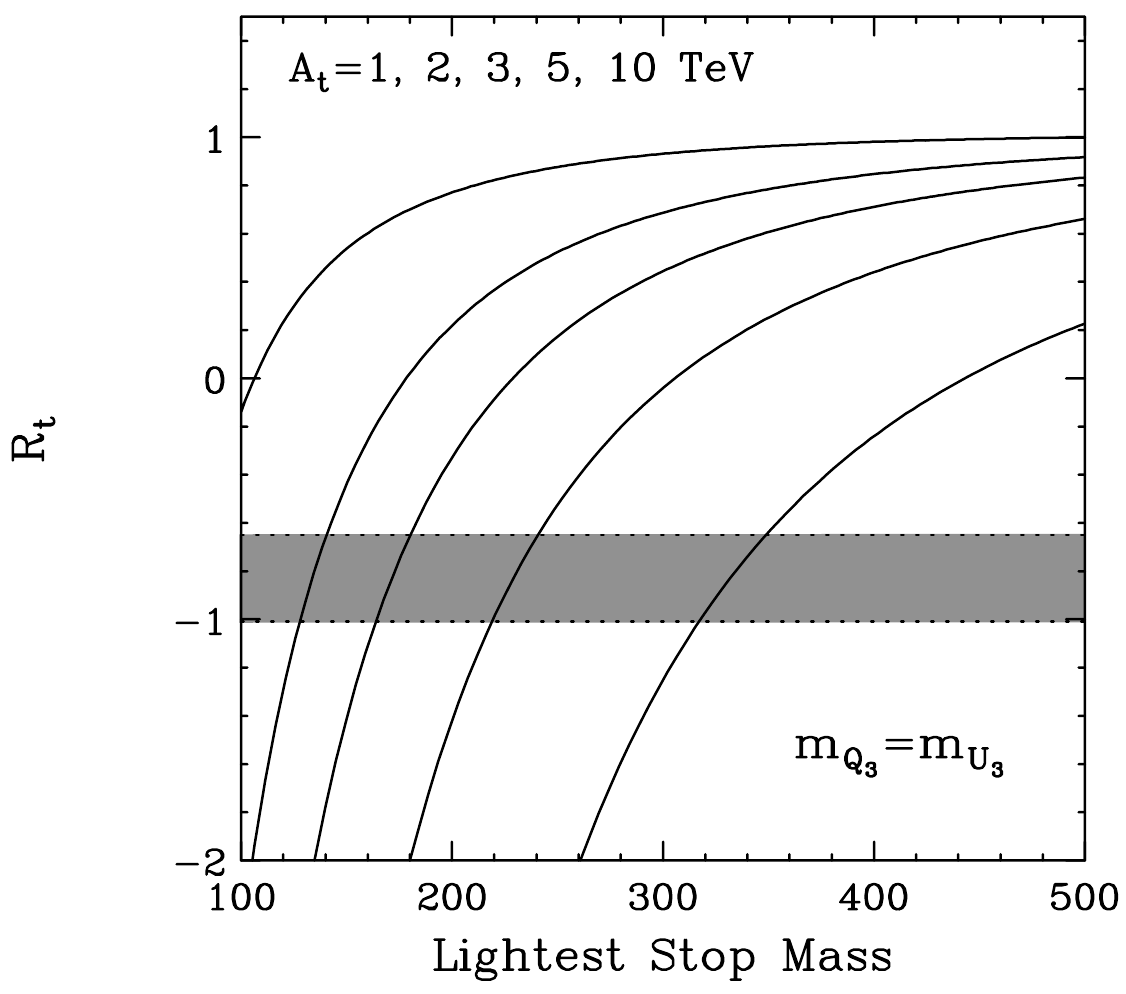}
\caption{$R_t$ as a function of the mass of the lightest stop, for several values of $A_t$. Here, we have assumed that $A_t \gg \mu \cot \beta$ and $m_{Q_3}=m_{U_3}$. The shaded horizontal band represents the range of $R_t$ that is favored by the global fit, as shown in Fig.~\ref{Rt}. For values of $A_t \gsim 2$ TeV and $m_{\tilde{t}_1} \lsim 300$ GeV, the favored range of $R_t$ can be accommodated.}
\label{stops}
\end{figure}

In Fig.~\ref{stops}, we show the value of $R_t$ as a function of the mass of the lightest stop, for several values of $A_t$. Here, we have assumed that $A_t \gg \mu \cot \beta$ and that the parameters $m_{Q_3}$ and $m_{U_3}$, appearing in the diagonal entries of the stop mass matrix, are equal to each other. We find that for values of $A_t \gsim 2$ TeV and $m_{\tilde{t}_1} \lsim 300$ GeV, values of $R_t$ consistent with those favored by the global fit to the Higgs data can be accommodated. While stops below 103.5~GeV are excluded by LEP-II~\cite{LEPstops}, the current searches from ATLAS~\cite{ATLASstops} and CMS~\cite{CMSstops} do not yet cover the full range of light stops, especially in the regime where the stop, top, and missing energy particle are nearly degenerate.

Such light and highly mixed stops may lead to problems with charge and color-breaking vacua. An approximate condition for metastability is sometimes given as \cite{Kusenko:1996jn}
\begin{equation}
A_t^2 +3\mu^2 \lesssim 7.5 (m_{\tilde{t}_1}^2+m_{\tilde{t}_2}^2).
\end{equation}
For the values of $R_t$ favored in our analysis, this condition is only met for stops lighter than the top. One could also consider models with particle content beyond the MSSM that may modify this requirement (see also Ref.~\cite{Reece:2012gi}).

\section{Conclusions}
\label{conclusions}

The discovery of the Higgs boson by the ATLAS and CMS experiments ushers in an exciting and much anticipated era in particle physics. The observed production mechanisms and decays (especially to pairs of electroweak gauge bosons) are sufficient to state, with reasonable certainty, that this new particle is intimately tied to the mechanism of electroweak symmetry breaking. The next task at hand is to ascertain whether the characteristics of this Higgs are consistent with those predicted by the Standard Model.

By considering the results of all reported Higgs channels from the LHC and Tevatron detector collaborations, we find that the observed rates of Higgs production followed by decays into $WW$, $ZZ$, and $\tau\tau$ are uniformly low compared to the Standard Model expectation. Furthermore, the rates observed in Higgs channels ending in $\gamma\gamma$ final states are uniformly high. This upward deviation is more pronounced after the application of selection criteria designed to isolate vector boson fusion and associated production diagrams over gluon-gluon fusion.

To accommodate this combination of observed rates, our fit favors a Higgs decay width to photons that is enhanced by a factor of approximately three, and a width to gluons that is suppressed by a factor of two, relative to the predictions of the Standard Model. As the widths of the Higgs to photons and gluons are set by loop-level interactions, these can be particularly sensitive to the presence of physics beyond the Standard Model. And although we must be cautious not to over-interpret these results, at face value, assuming no unknown systematics, we find that the best fit values for the Higgs widths to photons and gluons is preferred over those predicted by the Standard Model at greater than 99\% confidence. In contrast, our global fit finds no significant preference for modified widths to any of the Higgs' tree-level decay widths, such as $WW$, $ZZ$, $\tau\tau$, or $b\bar{b}$.

To alter the loop-level decays in a way that significantly improves the global fit, we postulate new particles with large couplings to the Higgs and that are charged under both QCD and electromagnetism. The most obvious candidates are partners of the top partner, such as stops within the context of supersymmetry. We find that the presence of such a particle can easily modify the Standard Model Higgs widths in such a way to come within approximately one standard deviation of the observed values. More specifically, we find that good fits to the data can be obtained for light ($\lesssim 300$~GeV) and well-mixed ($A_t \gsim 2$ TeV) stops. 

%Adding additional charged particles, such as staus or charginos, could further boost $\Gamma(h\to \gamma\gamma)$ without modifying the Higgs width to gluons, but are not statistically preferred at this time.

The full $\sim 20$~fb$^{-1}$ data set that is anticipated from the LHC by the end of 2012 will greatly reduce the experimental errors associated with the Higgs widths that drive our global fit. Furthermore, future consideration of systematic effects could plausibly modify the results under discussion here. With these caveats in mind, however, it appears that the presence the experimental evidence favors a Standard Model Higgs boson with modified decay widths to photons and gluons, suggesting the presence of new, strongly coupled physics present well below the TeV scale.

\section*{Aknowledgements}

The authors would like to thank Patrick Fox, Graham Kribs, Joseph Lykken, Tilman Plehn, Nausheen Shah, Alessandro Strumia, Carlos Wagner, and Marcela Carena for their useful advice on this project.
MRB and DH are supported by the US Department of Energy.

\end{document}